\documentclass[10pt, conference]{IEEEtran}
\IEEEoverridecommandlockouts

\usepackage{cite}
\usepackage{amsmath,amssymb,amsfonts}
\usepackage{algorithmic}
\usepackage{graphicx}
\usepackage{textcomp}
\usepackage{xcolor}
\def\BibTeX{{\rm B\kern-.05em{\sc i\kern-.025em b}\kern-.08em
    T\kern-.1667em\lower.7ex\hbox{E}\kern-.125emX}}

\usepackage{cite}
\usepackage{amsmath,amssymb,amsfonts}
\usepackage{algorithm}
\usepackage{algorithmic}
\usepackage{graphicx}
\usepackage{textcomp}
\usepackage{xcolor}
\usepackage{comment}
\usepackage{subcaption}
\usepackage{tikz}
\usepackage{pgfplots}
\pgfplotsset{compat=1.17}
\usepackage{amsmath}
\usepackage{multicol}
\usepackage{booktabs} 
\usepackage{multirow} 
\usepackage{adjustbox}
\usepackage{tcolorbox}
\usepackage{float}

\title{ALMAS: an Autonomous LLM-based Multi-Agent Software Engineering Framework}

\author
{\IEEEauthorblockN{Vali Tawosi*}
\IEEEauthorblockA{
\textit{J.P. Morgan AI Research}\\
London, UK \\
vali.tawosi@jpmorgan.com}
\and
\IEEEauthorblockN{Keshav Ramani*}
\IEEEauthorblockA{
\textit{J.P. Morgan AI Research}\\
New York, USA \\
keshav.ramani@jpmchase.com}
\and
\IEEEauthorblockN{Salwa Alamir*}
\IEEEauthorblockA{
\textit{J.P. Morgan AI Research}\\
London, UK \\
salwa.alamir@jpmchase.com}
\and
\IEEEauthorblockN{Xiaomo Liu}
\IEEEauthorblockA{
\textit{J.P. Morgan AI Research}\\
New York, USA \\
xiaomo.liu@jpmchase.com}
}

\begin{document}

\maketitle

\begin{abstract}
    
    Multi-agent Large Language Model (LLM) systems have been leading the way in applied LLM research across a number of fields. One notable area is software development, where researchers have advanced the automation of code implementation, code testing, code maintenance, \emph{inter alia}, using LLM agents. However, software development is a multifaceted environment that extends beyond just code. As such, a successful LLM system must factor in multiple stages of the software development life-cycle (SDLC).
    In this paper, we propose a vision for \textbf{ALMAS}, an \textbf{A}utonomous \textbf{L}LM-based \textbf{M}ulti-\textbf{A}gent \textbf{S}oftware Engineering framework, which follows the above SDLC philosophy such that it may work within an agile software development team to perform several tasks end-to-end.
    ALMAS aligns its agents with agile roles, and can be used in a modular fashion to seamlessly integrate with human developers and their development environment.
    We showcase the progress towards ALMAS through our published works and a use case demonstrating the framework, where ALMAS is able to seamlessly generate an application and add a new feature.
    
\end{abstract}


\section{Introduction}

Software development has evolved dramatically in recent years with the emergence of AI-assisted coding tools.
Although these tools have shown promise in tasks such as code completion, bug detection, maintenance \cite{alamir2022ai}, and documentation generation, they typically operate as isolated components rather than as an integrated ecosystem spanning the entire software development lifecycle. This fragmentation limits overall effectiveness and may introduce friction in developer workflows.

We introduce ALMAS (\textbf{A}utonomous \textbf{L}LM-based \textbf{M}ulti-\textbf{A}gent \textbf{S}oftware Engineer), a novel framework that orchestrates coding agents aligned with the diverse roles found in agile \cite{shore2021artofagile}, human-centric development teams: from product managers and sprint planners to developers, testers, and peer reviewers. By mirroring real-world team hierarchies, ALMAS deploys lightweight agents for routine, low-complexity tasks while assigning more advanced agents to handle complex architectural and integration decisions. This tiered approach not only aligns with the way human expertise is allocated in practice, but also ensures optimal resource utilization across development.

A key innovation of ALMAS lies in its dual operational modes that support both autonomous execution and interactive collaboration with human developers. This “three Cs” approach—Context-aware, Collaborative, and Cost-effective—ensures that specialized agents seamlessly communicate with one another as well as with their human teammates. As a result, the framework reduces cognitive load, enhances productivity, and promotes the cost-effective allocation of development resources.

The automation of the software development lifecycle (SDLC) has been a central focus where LLM-based multi-agent systems have emerged as effective solutions \cite{park2023generative}. Prior work in code generation \cite{yang2024sweagent}, computer control \cite{packer2023memgpt}, and web navigation \cite{zhou2023webarena} demonstrates the benefits of defining distinct agent roles with modular goals. Drawing on these insights, ALMAS leverages such modularity while addressing two common limitations of LLMs effectively: (1) context window length restrictions and (2) the diminishing effects of attention mechanisms for long prompts—owing to novel components that enable a compact natural language representation of codebases and a retrieval strategy that allows the LLM to effectively act as its own retriever for planning and execution.

Designed with industrial use in mind, ALMAS is envisioned to operate autonomously while seamlessly collaborating with human developers, ensuring smooth integration into real-world workflows. ALMAS agents were evaluated independently in previous works, but in this paper, we outline the framework's blueprint; detailing agent roles, interaction dynamics, and resource allocation strategies, and present a case study where an initial prototype successfully tackled a task that involved both the creation of a new application and the modification of existing code to add a new feature. ALMAS rapidly completed the task while integrating with common developer tools such as Atlassian Jira\footnote{https://www.atlassian.com/software/jira} and Bitbucket\footnote{https://bitbucket.org/product}.

In summary, ALMAS marks a significant evolution toward an end-to-end ecosystem for AI-assisted software engineering. By aligning agent roles with agile team dynamics and strategically allocating resources based on task complexity, the framework paves the way for an integrated, cost-effective, and context-aware automation of the software development lifecycle.

\begin{figure*}[ht]
  \centering
  \includegraphics[width=0.95\textwidth, trim={0 7.6cm 3cm 0.35cm},clip]{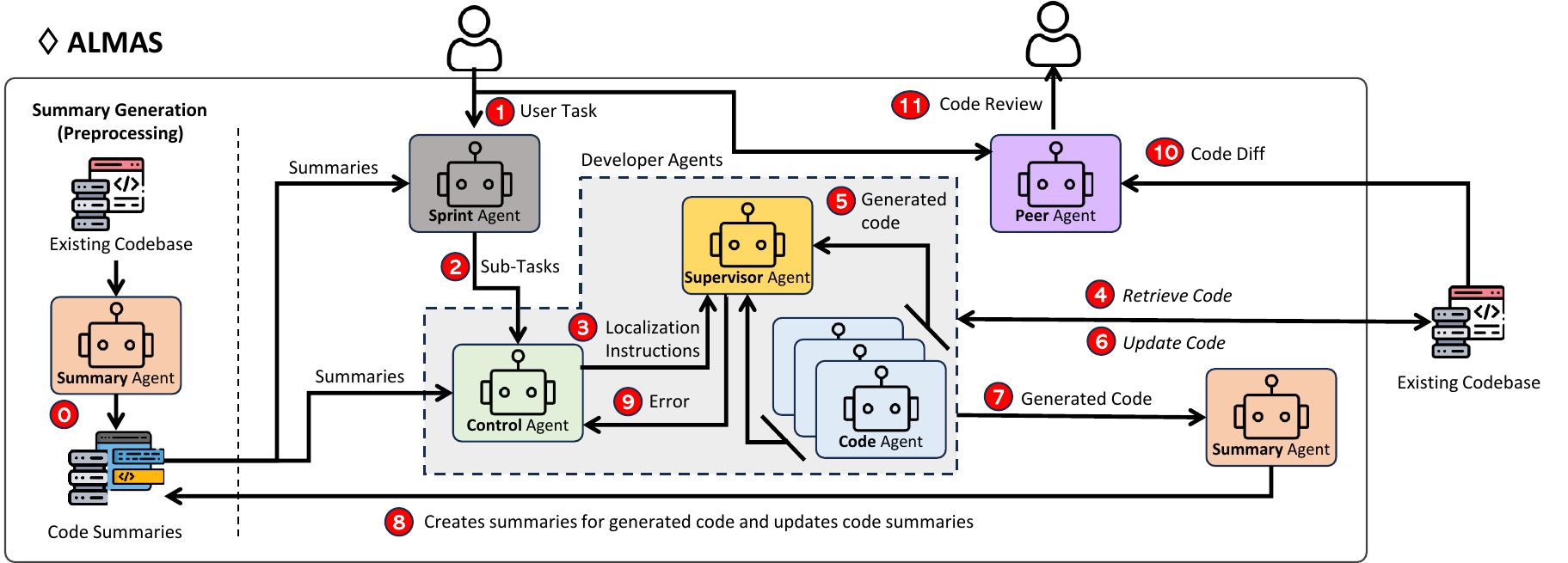}
  \caption{Overall conceptual system architecture diagram for ALMAS framework. }
  \label{fig:ALMAS-Arch}
\end{figure*}

\section{Related Work}

Large Language Models (LLMs) have been widely adopted in software engineering for tasks such as code summarization \cite{nam2024using, sun2024source}, program synthesis \cite{chen_2021evaluating}, code translation \cite{eniser2024towards}, automated repair \cite{xia2023automated, bouzenia2024repairagent}, and test generation \cite{ryan2024code}. Despite this focus on coding, only a fraction of developer effort is spent on implementation \cite{meyer_2021}. Motivated by this and recent surveys highlighting both the potential and current gaps of LLM-based tools \cite{he2024vision, li2024vicinagearth}, our work proposes that automation across the full SDLC can yield greater benefits. The ALMAS framework is designed to address this by supporting multiple SDLC phases in a multi-agent context.

\paragraph{Surveys and Overviews}
Recent surveys \cite{he2024vision, li2024vicinagearth} review LLM-driven multi-agent systems, discussing workflows, infrastructure, and challenges like context management. However, they stop short of proposing concrete solutions for issues such as task verification failures or agent misalignment.

\paragraph{Cognitive and Architectural Approaches}
Other studies have explored cognitive models for developer support \cite{leung2023} and modular agent architectures \cite{becattini2025}, but these typically address only specific SDLC stages. ALMAS extends these ideas, offering an end-to-end framework that manages agile role decomposition, dynamic orchestration, and context limitations.

\paragraph{Failure Analysis in Multi-Agent Systems}
Work by \cite{cemri2025} analyzes common failures in LLM-based multi-agent systems, such as poor task verification and communication breakdowns. ALMAS addresses these with agile role alignment, dynamic summarization, and a novel retrieval strategy to reduce such failures.

\paragraph{Code Augmentation and Retrieval}
Recent methods have enhanced code generation with retrieval mechanisms—Agentless \cite{xia2024agentless}, SWE-Agent \cite{yang2024sweagent}, and OpenHands \cite{wang2024openhands}—to improve bug fixing. Yet, these often overlook broader issues like context-window limits in large codebases. ALMAS advances this area by combining a new retrieval approach (Meta-RAG) with dynamic code summaries, supporting both bug fixes and feature development \cite{tawosi2025metaraglargecodebasesusing}.

In summary, while prior work has advanced individual aspects of LLM-based software engineering, most solutions are fragmented. ALMAS unifies these advances, coordinating agents for planning \cite{tawosi_2023}, localization \cite{tawosi2025metaraglargecodebasesusing}, generation, and review \cite{jensen_2024, agarwal2025codemiragehallucinationscodegenerated}, to address persistent challenges like limited context, attention dilution, and agent misalignment.

\section{Vision}

\begin{figure*}[th]
  \centering
  \includegraphics[width=0.95\textwidth, trim={0 6cm 0 0},clip]{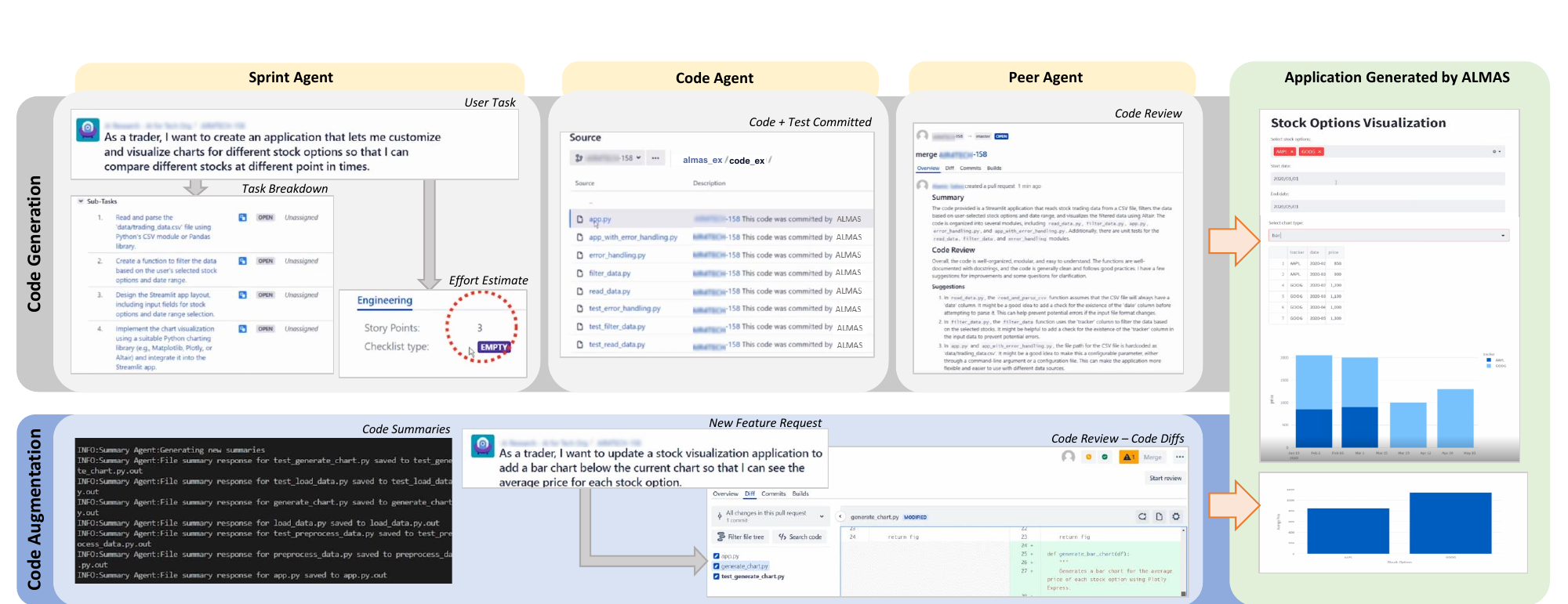}
  \caption{Example of application generated with ALMAS from a single task }
  \label{fig:ALMAS-Demo}
\end{figure*}

\textbf{Software Planning}: ALMAS begins with the Sprint Agent, acting as Product Manager and Scrum Master. Since unclear requirements can increase development risks \cite{letier2014uncertainty, hussain2016role}, the Sprint Agent refines user tasks for clarity and completeness, then breaks them into sub-tasks with descriptions, acceptance criteria, and effort estimates. Acceptance criteria support unit testing and code review, while effort estimation leverages few-shot learning from past examples \cite{tawosi_2023}. The Supervisor Agent allocates sub-tasks to the most suitable LLMs, optimizing for cost and performance by maintaining a diverse agent pool \cite{ong2024routellm}.

\textbf{Context-Aware Development}: The Summary and Control Agents enable context-aware development by preprocessing code repositories. The Summary Agent generates concise, structured natural-language summaries for each code unit, addressing LLM context limitations and reducing token costs. These summaries provide language-agnostic context for other agents. The Control Agent uses Retrieval Augmented Generation (RAG) over these summaries to localize relevant code for each sub-task, a process called Meta-RAG we introduced in \cite{tawosi2025metaraglargecodebasesusing}. The Sprint Agent also uses these summaries for task decomposition.

\textbf{Collaborative Development}: The Developer Agent coordinates multiple specialized agents to implement sub-tasks, using localized code and summaries. This collaborative setup allows agents to work alongside human developers, who can selectively integrate agents like Sprint or Peer Review into their workflow. The modular design supports easy replacement or ensembling of code generation solutions, enhancing flexibility and productivity.

\textbf{Cost Efficiency}: ALMAS is designed with cost efficiency in mind, leveraging several strategies to reduce cost without sacrificing performance. By condensing the codebase into structured natural-language replicas and keeping it up to date with the interim changes, the framework reduces token usage in the long run, which is a significant factor in LLM cost \cite{tawosi2025metaraglargecodebasesusing}.
The Supervisor Agent enhances cost efficiency by strategically routing tasks to the most suitable LLMs, considering their specialty, size, and cost \cite{ong2024routellm}. By having access to a diverse set of LLMs, it can select the optimal option for each task, optimizing resource use. This modular design allows teams to customize the framework to their needs, further reducing costs through selective agent utilization.

\textbf{Validation, Verification, and Error Handling}: Validation is built in: the Developer Agent checks code formatting, compilation, and runs unit tests to ensure correctness. The Peer Agent reviews code for functionality, vulnerabilities \cite{jensen_2024}, performance, hallucinations \cite{agarwal2025codemiragehallucinationscodegenerated}, formal verification \cite{ramani2025formalmethod}, and quality, providing a report for human review. For error handling, failed tests trigger the Control Agent to localize and address issues, while the Supervisor Agent tracks agent actions. If automated recovery fails after several attempts, control passes to a human developer with a summarized action history, ensuring resilience.

\textbf{Tool Usage and Integration}: ALMAS integrates seamlessly with SDLC tools—CI/CD, version control, task management—and can connect to IDEs like VS Code or IntelliJ via plugins. This ensures it fits naturally into existing development workflows, serving as both a standalone and complementary solution.

\section{Progress Towards Vision}

A number of ALMAS agents have been developed, including Summary Agent, Control Agent, Sprint Agent, one Code Agent, and Peer Agent. The ALMAS framework aims to join these into a multi-agent system that aligns with real-world software engineering roles. In order to demonstrate the potential of the wider vision, these agents were tasked with the creation of a Python streamlit application. 
All agents use GPT-4o for this exercise. 
The agents were also given access to interact with Atlassian tools; Jira for task management and Bitbucket for code versioning. Figure \ref{fig:ALMAS-Demo} illustrates the automated software development workflow utilizing the multi-agent system architecture.

The workflow is divided into two main phases: Code Generation and Code Augmentation. In the Code Generation phase, the Sprint Agent initiates the process by interpreting user requirements and breaking them down into manageable sub-tasks, providing effort estimates in the form of story points. The Code Agent then develops the application by writing and committing code, and corresponding unit tests. To ensure code quality, the Peer Agent conducts thorough code reviews, offering feedback and recommendations for improvement. The final output of this phase is the automated generation of the application by ALMAS, resulting in a stock options visualization tool. 

The Code Augmentation phase starts with generating code summaries and supports iterative development through new feature requests, such as adding a bar chart for average stock prices. The Control Agent identifies necessary code snippets for the Code Agent to make progress towards fulfilling the user task. The Code Agent, then, generates and commits code and unit tests.
The Peer Agent revisits the code review process, and in this phase we focus on code differences to demonstrate implementation of new features within the existing codebase. Finally, the application is updated, incorporating the new feature to enhance its functionality and user experience. 

This multi-agent system architecture creates a modular framework by which the agents can be used individually or in combination. As such, they can be configured such that each agent can use a different LLM, taking into account their need for a specialized LLM or LLMs of different sizes and costs. Furthermore, this design has allowed us to conduct extensive research into each agent in isolation. 
Nevertheless, this paper demonstrates the wider vision of the interaction of all specialized agents towards one common purpose. The necessary end-to-end evaluation will be explored more thoroughly in future.

\section{Conclusion and Future Work}

In this paper, we presented ALMAS—a multi-agent LLM-based framework that embodies the diverse roles of an agile software development team. By integrating specialized agents for sprint planning, code generation, and code review, \textit{inter alia}, ALMAS effectively automates multiple stages of the software development lifecycle. Our illustrative example, where ALMAS successfully built a Streamlit application and later augmented it with a new feature, demonstrates the framework’s potential for automated software development.

Looking ahead, we plan to conduct end-to-end evaluations of ALMAS on a range of coding tasks. For instance, using benchmarking datasets like SWE-Bench, we aim not only to assess its capability in resolving bug fixes but also to measure intermediate metrics such as localization efficiency.

\section*{Disclaimer}
This paper was prepared for informational purposes by the Artificial Intelligence Research group of JPMorgan Chase \& Co and its affiliates (“JP Morgan”), and is not a product of the Research Department of JP Morgan. JP Morgan makes no representation and warranty whatsoever and disclaims all liability, for the completeness, accuracy or reliability of the information contained herein. This document is not intended as investment research or investment advice, or a recommendation, offer or solicitation for the purchase or sale of any security, financial instrument, financial product or service, or to be used in any way for evaluating the merits of participating in any transaction, and shall not constitute a solicitation under any jurisdiction or to any person, if such solicitation under such jurisdiction or to such person would be unlawful.

\bibliographystyle{IEEEtran}
\bibliography{ase2025/references}

\begin{thebibliography}{10}
\providecommand{\url}[1]{#1}
\csname url@samestyle\endcsname
\providecommand{\newblock}{\relax}
\providecommand{\bibinfo}[2]{#2}
\providecommand{\BIBentrySTDinterwordspacing}{\spaceskip=0pt\relax}
\providecommand{\BIBentryALTinterwordstretchfactor}{4}
\providecommand{\BIBentryALTinterwordspacing}{\spaceskip=\fontdimen2\font plus
\BIBentryALTinterwordstretchfactor\fontdimen3\font minus \fontdimen4\font\relax}
\providecommand{\BIBforeignlanguage}[2]{{%
\expandafter\ifx\csname l@#1\endcsname\relax
\typeout{** WARNING: IEEEtran.bst: No hyphenation pattern has been}%
\typeout{** loaded for the language `#1'. Using the pattern for}%
\typeout{** the default language instead.}%
\else
\language=\csname l@#1\endcsname
\fi
#2}}
\providecommand{\BIBdecl}{\relax}
\BIBdecl

\bibitem{alamir2022ai}
S.~Alamir, P.~Babkin, N.~Navarro, and S.~Shah, ``{AI} for automated code updates,'' in \emph{Proceedings of the 44th International Conference on Software Engineering: Software Engineering in Practice}, 2022, pp. 25--26.

\bibitem{shore2021artofagile}
J.~Shore and S.~Warden, \emph{The art of agile development}.\hskip 1em plus 0.5em minus 0.4em\relax " O'Reilly Media, Inc.", 2021.

\bibitem{park2023generative}
J.~S. Park, J.~O'Brien, C.~J. Cai, M.~R. Morris, P.~Liang, and M.~S. Bernstein, ``Generative agents: Interactive simulacra of human behavior,'' in \emph{Proceedings of the 36th annual acm symposium on user interface software and technology}, 2023, pp. 1--22.

\bibitem{yang2024sweagent}
J.~Yang, C.~E. Jimenez, A.~Wettig, K.~Lieret, S.~Yao, K.~Narasimhan, and O.~Press, ``Swe-agent: Agent-computer interfaces enable automated software engineering,'' \emph{arXiv preprint arXiv:2405.15793}, 2024.

\bibitem{packer2023memgpt}
C.~Packer, V.~Fang, S.~G. Patil, K.~Lin, S.~Wooders, and J.~E. Gonzalez, ``Memgpt: Towards llms as operating systems,'' \emph{arXiv preprint arXiv:2310.08560}, 2023.

\bibitem{zhou2023webarena}
S.~Zhou, F.~F. Xu, H.~Zhu, X.~Zhou, R.~Lo, A.~Sridhar, X.~Cheng, T.~Ou, Y.~Bisk, D.~Fried \emph{et~al.}, ``Webarena: A realistic web environment for building autonomous agents,'' \emph{arXiv preprint arXiv:2307.13854}, 2023.

\bibitem{nam2024using}
D.~Nam, A.~Macvean, V.~Hellendoorn, B.~Vasilescu, and B.~Myers, ``Using an llm to help with code understanding,'' in \emph{Proceedings of the IEEE/ACM 46th International Conference on Software Engineering}, 2024, pp. 1--13.

\bibitem{sun2024source}
W.~Sun, Y.~Miao, Y.~Li, H.~Zhang, C.~Fang, Y.~Liu, G.~Deng, Y.~Liu, and Z.~Chen, ``Source code summarization in the era of large language models,'' \emph{arXiv preprint arXiv:2407.07959}, 2024.

\bibitem{chen_2021evaluating}
M.~C. et~al., ``Evaluating large language models trained on code,'' 2021.

\bibitem{eniser2024towards}
H.~F. Eniser, H.~Zhang, C.~David, M.~Wang, B.~Paulsen, J.~Dodds, and D.~Kroening, ``Towards translating real-world code with llms: A study of translating to rust,'' \emph{arXiv preprint arXiv:2405.11514}, 2024.

\bibitem{xia2023automated}
C.~S. Xia, Y.~Wei, and L.~Zhang, ``Automated program repair in the era of large pre-trained language models,'' in \emph{2023 IEEE/ACM 45th International Conference on Software Engineering (ICSE)}.\hskip 1em plus 0.5em minus 0.4em\relax IEEE, 2023, pp. 1482--1494.

\bibitem{bouzenia2024repairagent}
I.~Bouzenia, P.~Devanbu, and M.~Pradel, ``Repairagent: An autonomous, llm-based agent for program repair,'' \emph{arXiv preprint arXiv:2403.17134}, 2024.

\bibitem{ryan2024code}
G.~Ryan, S.~Jain, M.~Shang, S.~Wang, X.~Ma, M.~K. Ramanathan, and B.~Ray, ``Code-aware prompting: A study of coverage-guided test generation in regression setting using llm,'' \emph{Proceedings of the ACM on Software Engineering}, vol.~1, no. FSE, pp. 951--971, 2024.

\bibitem{meyer_2021}
A.~N. Meyer, E.~T. Barr, C.~Bird, and T.~Zimmermann, ``Today was a good day: The daily life of software developers,'' \emph{IEEE Transactions on Software Engineering}, vol.~47, no.~05, pp. 863--880, may 2021.

\bibitem{he2024vision}
\BIBentryALTinterwordspacing
J.~He, C.~Treude, and D.~Lo, ``Llm-based multi-agent systems for software engineering: Literature review, vision and the road ahead,'' 2024. [Online]. Available: \url{https://arxiv.org/abs/2404.04834}
\BIBentrySTDinterwordspacing

\bibitem{li2024vicinagearth}
\BIBentryALTinterwordspacing
X.~Li, S.~Wang, S.~Zeng, Y.~Wu, and Y.~Yang, ``A survey on llm-based multi-agent systems: workflow, infrastructure, and challenges,'' \emph{Vicinagearth}, vol.~1, no.~1, p.~9, Oct 2024. [Online]. Available: \url{https://doi.org/10.1007/s44336-024-00009-2}
\BIBentrySTDinterwordspacing

\bibitem{leung2023}
M.~Leung and G.~Murphy, ``On automated assistants for software development: The role of llms,'' in \emph{2023 38th IEEE/ACM International Conference on Automated Software Engineering (ASE)}, 2023, pp. 1737--1741.

\bibitem{becattini2025}
M.~Becattini, R.~Verdecchia, and E.~Vicario, ``Sallma: A software architecture for llm-based multi-agent systems,'' in \emph{2025 IEEE/ACM International Workshop New Trends in Software Architecture (SATrends)}, 2025, pp. 5--8.

\bibitem{cemri2025}
\BIBentryALTinterwordspacing
M.~Cemri, M.~Z. Pan, S.~Yang, L.~A. Agrawal, B.~Chopra, R.~Tiwari, K.~Keutzer, A.~Parameswaran, D.~Klein, K.~Ramchandran, M.~Zaharia, J.~E. Gonzalez, and I.~Stoica, ``Why do multi-agent llm systems fail?'' 2025. [Online]. Available: \url{https://arxiv.org/abs/2503.13657}
\BIBentrySTDinterwordspacing

\bibitem{xia2024agentless}
C.~S. Xia, Y.~Deng, S.~Dunn, and L.~Zhang, ``Agentless: Demystifying llm-based software engineering agents,'' \emph{arXiv preprint arXiv:2407.01489}, 2024.

\bibitem{wang2024openhands}
X.~Wang, B.~Li, Y.~Song, F.~F. Xu, X.~Tang, M.~Zhuge, J.~Pan, Y.~Song, B.~Li, J.~Singh \emph{et~al.}, ``Openhands: An open platform for ai software developers as generalist agents,'' \emph{arXiv preprint arXiv:2407.16741}, 2024.

\bibitem{tawosi2025metaraglargecodebasesusing}
\BIBentryALTinterwordspacing
V.~Tawosi, S.~Alamir, X.~Liu, and M.~Veloso, ``Meta-rag on large codebases using code summarization,'' 2025. [Online]. Available: \url{https://arxiv.org/abs/2508.02611}
\BIBentrySTDinterwordspacing

\bibitem{tawosi_2023}
\BIBentryALTinterwordspacing
V.~Tawosi, S.~Alamir, and X.~Liu, \emph{Search-Based Optimisation of LLM Learning Shots for Story Point Estimation}.\hskip 1em plus 0.5em minus 0.4em\relax Springer Nature Switzerland, Dec. 2023, p. 123–129. [Online]. Available: \url{http://dx.doi.org/10.1007/978-3-031-48796-5\_9}
\BIBentrySTDinterwordspacing

\bibitem{jensen_2024}
R.~I.~T. Jensen, V.~Tawosi, and S.~Alamir, ``Software vulnerability and functionality assessment using llms,'' 2024.

\bibitem{agarwal2025codemiragehallucinationscodegenerated}
\BIBentryALTinterwordspacing
V.~Agarwal, Y.~Pei, S.~Alamir, and X.~Liu, ``Codemirage: Hallucinations in code generated by large language models,'' 2025. [Online]. Available: \url{https://arxiv.org/abs/2408.08333}
\BIBentrySTDinterwordspacing

\bibitem{letier2014uncertainty}
E.~Letier, D.~Stefan, and E.~T. Barr, ``Uncertainty, risk, and information value in software requirements and architecture,'' in \emph{Proceedings of the 36th International Conference on Software Engineering}, 2014, pp. 883--894.

\bibitem{hussain2016role}
A.~Hussain, E.~O. Mkpojiogu, and F.~M. Kamal, ``The role of requirements in the success or failure of software projects,'' \emph{International Review of Management and Marketing}, vol.~6, no.~7, pp. 306--311, 2016.

\bibitem{ong2024routellm}
I.~Ong, A.~Almahairi, V.~Wu, W.-L. Chiang, T.~Wu, J.~E. Gonzalez, M.~W. Kadous, and I.~Stoica, ``Routellm: Learning to route llms from preference data,'' in \emph{The Thirteenth International Conference on Learning Representations}, 2024.

\bibitem{ramani2025formalmethod}
K.~Ramani, V.~Tawosi, S.~Alamir, and D.~Borrajo, ``Bridging llm planning agents and formal methods: A case study in plan verification,'' in \emph{Proceedings of the 1st International Workshop on Autonomous Agents in Software Engineering (AgenticSE)}, 2025.

\end{thebibliography}
\end{document}